\documentclass[usenatbib]{mnras}

\usepackage{graphicx}

\usepackage{times}

\title{Strongly magnetized accretion in two ultracompact binary systems}

\author[Maccarone et al.]
{Thomas J. Maccarone,$^{1\ast}$ Thomas Kupfer,$^{1,7}$ Edgar Najera~Casarrubias$^{1}$,
Liliana E. Rivera~Sandoval$^2$, 
\newauthor 
Aarran W. Shaw$^{3,8}$, Christoper T. Britt$^4$,
Jan van~Roestel$^5$ and David R. Zurek$^6$\\
\normalsize{$^{1}$Department of Physics \& Astronomy, Texas Tech University, Box 41051, Lubbock, TX, USA, 79409-1051}\\
\normalsize{$^{2}$Department of Physics \& Astronomy, University of Texas-Rio Grande Valley, Brownsville, TX, USA}\\
\normalsize{$^{3}$Department of Physics and Astronomy, Butler University, 4600 Sunset Ave, Indianapolis, IN, 46208, USA}\\
\normalsize{$^{4}$Space Telescope Science Institute, Baltimore, MD, USA}\\
\normalsize{$^{5}$Astronomical Institute `Anton Pannekoek', University of Amsterdam, Amsterdam, NL}\\
\normalsize{$^{6}$American Museum of Natural History, New York, NY, USA }\\
\normalsize{$^{7}$Hamburger Sternwarte, University of Hamburg, Gojenbergsweg 112, 21029 Hamburg, Germany}\\
\normalsize{$^{8}$Department of Physics, University of Nevada, Reno, 1664 N. Virginia Street, Reno, NV, 89557, USA}\\
\\
\normalsize{$^\ast$To whom correspondence should be addressed; E-mail:  thomas.maccarone@ttu.edu.}
}

\date{}

\begin{document} 



\maketitle


\begin{abstract}
 We present the discoveries of two of AM CVn systems,{ Gaia14aae and SDSS~J080449.49+161624.8, which show X-ray pulsations at their orbital periods, indicative of magnetically collimated accretion}.  { Both also show indications of higher rates of mass transfer relative to the expectations from binary evolution driven purely by gravitational radiation, based on existing optical data for Gaia14aae, which show a hotter white dwarf temperature than expected from standard evolutionary models, and X-ray data for SDSS~J080449.49+161624.8 which show a luminosity 10-100 times higher than those for other AM~CVn at similar orbital periods. The higher mass transfer rates could be driven by magnetic braking from the disk wind interacting with the magnetosphere of the tidally locked accretor.  We discuss implications of this additional angular momentum transport mechanism for evolution and gravitational wave detectability of AM CVn objects.} 
\end{abstract}


\section{Introduction}

Double compact object binaries are one of the primary sources of gravitational waves in the universe.  The subset with two white dwarfs is expected to dominate the sky for the Laser Interferometer Space Antenna \citep[LISA;][]{LISA},  which will take advantage of the absence of seismic noise in space to detect gravitational waves at frequencies from $10^{-4}$ Hz to 1 Hz \citep{LISA}. The LISA verification binaries are objects known, and electromagnetically characterized, at the time of LISA's launch, and these include substantial numbers of double white dwarf binaries, both with and without mass transfer between the two objects \citep{LISA}.

The evolution of close binary stars is driven by three key processes:  angular momentum transport (typically from gravitational radiation and/or magnetic braking) \citep{1962ApJ...136..312K, 1981A&A...100L...7V}; mass transfer (both between components of the binary and expulsion of material from the binary in winds); and evolution of the stars in the binaries themselves (see \citealt{PostnovYungelson} for a review).  

In the AM CVn binaries, the prevailing assumptions in nearly all theory work are that the angular momentum transport out of the system is solely due to gravitational radiation and that the mass transfer is conservative (e.g. \citealt{Nelemans2001}).  

Still, it has been shown that the standard tracks for AM~CVn systems can fail to describe their evolution accurately if there are magnetic fields of $10^3$G or more for the accretor white dwarfs, because the disk winds from the accretion disk, in combination with the magnetic field of the white dwarf, will drive magnetic braking angular momentum loss, which could exceed the angular momentum loss from gravitational radiation \citep{2010arXiv1006.4112F}.  This effect should become more important for either higher magnetic fields (since the magnetic braking torques depend on the magnetic field) or longer orbital periods (since the gravitational wave angular momentum losses are slower for longer periods), and magnetic fields of $10^3$ G are not exceptionally high for white dwarfs.

Traditionally, it has been hard to find evidence for such magnetic fields in the AM CVn systems.  One system, SDSS~J080449.49+161624.8 (SDSS~J0804), shows a single-peaked helium emission line, something which is often indicative of magnetically dominated accretion in white dwarfs with hydrogen-rich donor stars \citep{Roelofs2009}, but clear evidence for magnetic accretion must come in the form of a measurement of modulation of the accretion power on the accretor's spin period.  In this paper, we show evidence from two sources' X-ray pulsations on the orbital period that they have strong magnetic fields.  Additionally, we show that one of them clearly has a mass transfer rate well in excess of that expected from standard binary evolution theory, providing strong evidence that the enhanced magnetic braking from the accretor is strongly affecting its evolution.

\section{Observations and data reduction}

\begin{figure*}%
\centering
\includegraphics[width=0.4\textwidth]{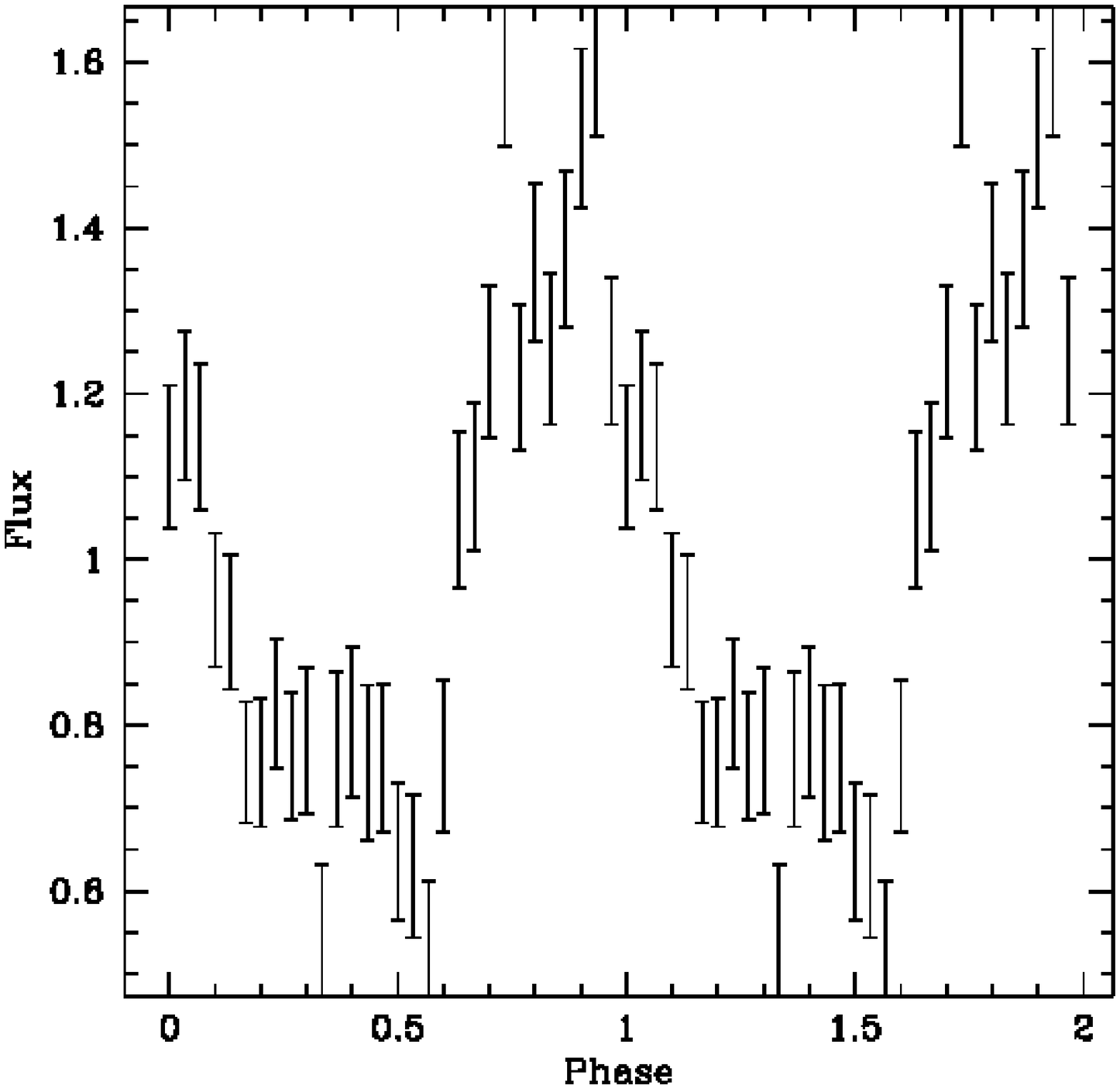}
\includegraphics[width=0.4\textwidth]{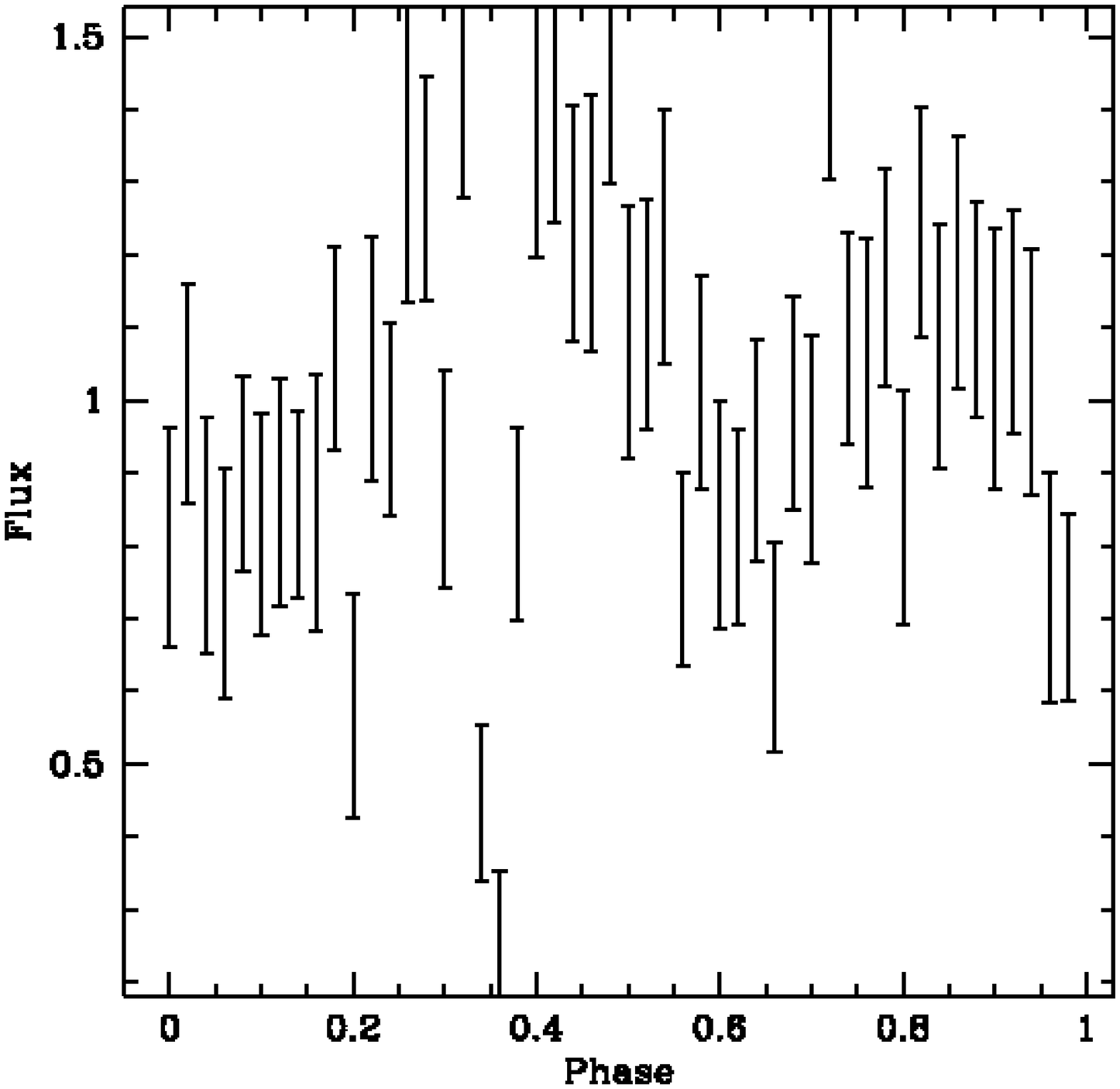}
\caption{Left: The folded light curve for SDSS~J0804 at the 44.5 min period, normalized as a ratio to the mean count rate.  Error bars are 1$\sigma$.  The ephemeris for the folding is set arbitrarily, as it is not well enough known to match.  Right: The folded light curve for Gaia14aae at the 49.7 min period, normalized as a ratio to the mean count rate.  Error bars are 1$\sigma.$  The ephemeris for folding is set arbitrarily, as it is not well enough known to match, and the X-ray eclipse lines up properly with that seen by the XMM-Newton Optical Monitor. 
 { In both plots, the fluxes plotted are ratios to the average flux during the observation.}}\label{folded_lcs}
\end{figure*}

SDSS~J080449.49+161624.8 (SDSS~J0804) was observed by XMM-Newton on 16 April 2018 from 14:28:45 to 17:47:22.  For this source, the XMM-Newton GOF has produced standard pipeline light curves with the three detectors summed.  We used these light curves, which are from 0.2 to 10 keV, and fold them on the known 44.5 minute orbital period \citep{Roelofs2009}.  These results are shown in figure \ref{folded_lcs}.

Gaia14aae was observed by XMM-Newton on 12 January 2023 from 3:24:58 to 17:39:08.  The XMM-Newton data are analyzed using standard procedures.  After applying standard screening, we extracted light curves from each of the three X-ray cameras.  The results shown are for energies from 0.2-10 keV.  For each camera, we created an off-source background file, and used the evselect command in the XMMSAS software to produce light curves.  We then corrected for exposure duty cycle using lccorr.  We then added the three light curves using the FTOOLS task lcmath.  The light curves were originally produced with a time resolution of 7.8 seconds, corresponding to the readout time for the slower MOS detectors.  Following that, we folded the light curves using the FTOOLS efold task.  The folded light curves in figure \ref{folded_lcs} are folded on the orbital period of 49.7 minutes from optical measurements \citep{Campbell}.

%We show here, in figure \ref{0804_folded}, that this system shows strong modulation of its X-ray emission on the orbital period of 44.5 minutes found by \citep{Roelofs2009}.  This system shows a single-peaked X-ray pulsation, suggestive of truncation of the accretion disk relatively far from the white dwarf \citep{1999A&A...347..203N}, and hence suggestive of a relatively high magnetic field, of $\sim10^5$ G or more.

%We find a similar X-ray pulsation for another AM~CVn system Gaia14aae.  Gaia14aae is an eclipsing binary with an orbital period of 49.7 minutes \citep{Campbell}.  We fold the data from an XMM-Newton observation on the known orbital period.  The folded X-ray light curve is plotted in figure \ref{Gaia_folded} and shows a strong eclipse at the same orbital phase as the ultraviolet light curve obtained from the simultaneous Optical Monitor data from XMM-Newton.  On top of the variability from the eclipse, the system also shows a tell-tale double-peaked additional modulation.

\section{Analysis}

\subsection{X-ray fluxes and luminosities}
For Gaia14aae, the flux is found to be $3.1\times10^{-13}$ erg/sec/cm$^2$, from the XMM SAS standard pipeline analysis.  Taking the distance from Gaia EDR3 of $258\pm8$ pc \citep{GaiaEDR3,ramsay2018}, we find an X-ray luminosity of $2.4\times10^{30}$ erg/sec.  This value is bit lower than that for other AM CVn at similar orbital periods \citep{ramsay2006}, but X-ray emission may be suppressed due to inclination angle effects \citep{vanteesling}, as it is an eclipsing binary.  { The source's optical luminosity is similarly likely to be underestimated if one takes $L=4\pi{d^2}F$. \citet{Campbell} have found from the temperature of the accreting white dwarf that the accretion rate is about $7.5\times10^{-11} M_\odot$/yr.  If the system began mass transfer very shortly after the birth of the more massive white dwarf, and hence the temperature reflects the initial conditions for the system and not its quasi-steady state, this could represent an overestimate of the mass transfer rate, but it is also the case that the presence of frequent accretion disk outbursts in a system at such a long orbital period is indicative of a higher mass transfer rate than the $1.5\times10^{-12} M_\odot$/year expected \citep{Kotko2012,Campbell}.
}
{ For SDSS~J0804, there is not a good white dwarf temperature estimate, so its mass transfer rate must be estimated from its luminosity. The Gaia parallax distance estimate for SDSS~J0804 is  999$\pm^{185}_{134}$ pc \citep{BailerJones} yielding  $L_X$=$3\times10^{32}$ erg/sec, a { factor of 10-100} times brighter than the X-ray luminosities for other AM~CVn at similar orbital periods \citep{2023arXiv231206007B}.  We can take the sum of the X-ray and optical luminosities for the system, and estimate the mass transfer rate, $\dot{m}$ using:
\begin{equation}
    \dot{m}=\frac{2RL_X}{GM},
\end{equation}
where $R$ is the accreting white dwarf radius, $L_X$ is the X-ray luminosity, and $M$ is the accretor mass. 
 This assumes that the X-rays come from the emission in the boundary layer, and could be an overestimate by a factor of 2 if the X-rays dominate the total accretion flow emission due to a disk truncated far from the white dwarf.  This yields a mass transfer rate of about $9\times10^{-11} M_\odot$/yr, again, much higher than the predictions from standard binary evolution due to gravitational radiation alone.}

{ We note that such a high accretion rate could be consistent with expectations if the donor were a helium star of significantly higher mass than a white dwarf that fills its Roche lobe for the same period.  Given that the chemical composition of the donor inferred from spectroscopy is consistent with a white dwarf and inconsistent with a helium star \citep{Nelemans2010}, this scenario is unlikely to be relevant for SDSS~0804.}

\subsection{Evidence for magnetic accretion and estimation of the white dwarf magnetic fields}

From figure \ref{folded_lcs}, both objects clearly show X-ray emission modulated by the spin period.  For SDSS~J0804, the emission is single-peaked.  The orbital solution for this object is not well enough determined for us to determine the phasing of the X-ray peak with respect to the orbital phases.  For Gaia14aae, it is clear that the eclipse occurs slightly before the X-ray maximum, and the pulse profile is double-peaked.  In both systems, the maximum X-ray flux is about 1.5 times the mean X-ray flux.

For SDSS~J0804, the amplitude alone provides strong evidence that the periodicity is due to magnetically collimated accretion.  This system is known not to be eclipsing, and the X-ray emitting region of a boundary layer on the surface of the accretor is harder to eclipse than the much larger accretion disk around the white dwarf.  Modulations from a disk wind's differential absorption can sometimes yield periodicities in edge-on ultracompact X-ray binaries (e.g. \citealt{Bahramian2017}), but with very low amplitude.

For Gaia14aae, the system is at high inclination, and does eclipse.  Here phasing of the non-eclipse orbital modulation, with a strong peak roughly synchronous with the eclipse, removes the possibility of geometric modulations via variable absorption in a disk wind.  Such modulations should leave the lowest fluxes just outside the eclipse, not the highest ones, and they would also be stronger in the softer X-rays, where the strongest atomic absorption features are, but we find no strong energy dependence for the amplitude of modulation.  Modulations could show this phasing if they involved enhanced emission due to scattering off the hot spot where the accretion stream impacts the outer accretion disk, but such a mechanism could not produce the $\approx 50\%$ amplitude seen, and the double-peaked profile would also not result in this case.

The only viable mechanism for the modulations in both sources, then, is polar accretion.  In the approximation of spherical inflow, the Alfv\'en radius can be set equal to \citep{2002apa..book.....F} :
\begin{equation}
    r_A = 7.6\times 10^8 {\rm cm} \left(\frac{\dot{M}}{M_{f}}\right)^{-\frac{2}{7}} \left(\frac{B}{10^3 {\rm G}}\right)^{\frac{4}{7}} \left(\frac{R}{10^9 {\rm cm}}\right)^{-\frac{12}{7}} \left(\frac{M}{M_\odot} \right)^{-\frac{1}{7}},
\end{equation}
{ where $M_{f}$ is the fiducial accretion rate of $7.5\times10^{-11} M_\odot$/yr, based on estimates} of the accretion rate from \cite{Campbell} for Gaia14aae.  This gives us some bounds on the range of magnetic fields that is plausible for the system.  
\subsubsection{Magnetic field estimation for Gaia14aae}
The magnetic field strength must be large enough to disrupt the accretion disk outside the surface of the white dwarf (given that pulsations are seen), but also small enough to allow an outer disk to form given that outbursts are seen in both sources, and that the disc signature shows up in eclipse mapping \citep{Campbell}).

\begin{figure}
    \centering
    \includegraphics[angle=-90,width=3 in]{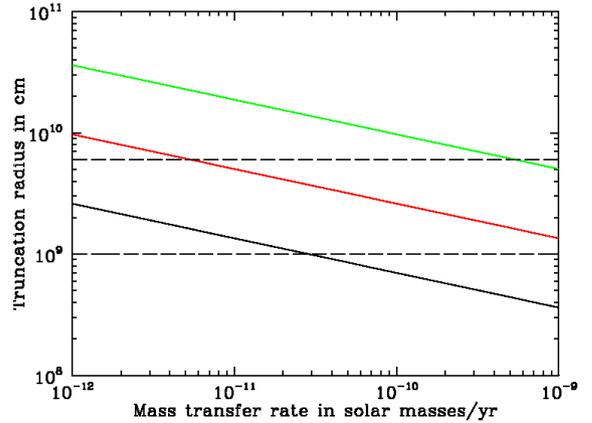}
    \caption{The truncation radius of an accretion disc plotted versus mass transfer rate.  The upper dashed line is the approximate circularization radius of an AM CVn with an orbital period in the 45-50 minute range.  The lower line is an approximate accretor white dwarf radius.  The black line represents a magnetic field of $10^3$ G, the red line $10^4$ G, and the green line $10^5$G.  We can see that for the range near $7-9\times 10^{-11} M_\odot$/yr, the red curve is comfortably between the two radii, yielding a truncated disc, while the green curve is above the circularization radius line, indicating that no disc can form.}
    \label{fig:alfven}
\end{figure}

The semi-major axis for the orbit of Gaia14aae should be about $3\times10^{10}$ cm, and the outer radius of the accretion disk is typically 0.2-0.3 times the orbital separation for relatively extreme mass ratio systems \citep{2002apa..book.....F}. The magnetic field of the white dwarf must then be in the range of about $10^3-10^5$ G in order for there to be a disk that forms, and is truncated (see Figure \ref{fig:alfven}).  Other classes of white dwarfs, including isolated white dwarfs, the accretors in cataclysmic variables \citep{Pala}, and the accretors in the symbiotic stars which are the likely progenitors of AM CVn systems \citep{Sokoloski} all show substantial subsets of the objects with magnetic fields of $10^5$ G or more.

We can gain some further, albeit qualitative, insights on the magnetic field strength in Gaia14aae from the presence of a double-peaked X-ray pulse profile.  The double-peaked pulse profile implies that the accretor has a relatively weak magnetic field, with the truncation of the thin accretion disk fairly close to the surface of the white dwarf \citep{1999A&A...347..203N}.  In principle, the single and double pulse peaks could result from accretion along one magnetic pole or both poles, respectively, but the systematic correlation between number of peaks and inferred magnetic field, and the other evidence for a stronger magnetic field in SDSS~0804 than in Gaia14aae both point to variations in the magnetic field, and in the truncation radii, as the determining factor for the pulse profile shape.  For the intermediate polar CVs with non-degenerate donor stars, this effect manifests itself as double-peaked pulse profiles showing up for the shorter spin period systems, in which the magnetic torques causing spin-down are weaker. For the ultracompact binaries, the spin period is set by the tidal locking \citep{tidal_locking} (and these results provide clearer observational evidence than has previously existed for the tidal locking), so it is not a tracer of the accretor's magnetic moment, but the relationship between double-peaked pulse profiles and truncation close to the white dwarf's surface should remain.  This, in turn, means that the magnetic field is likely to be in the range of $10^{3-4}$ G. 

\subsubsection{Magnetic field estimation for SDSS~J0804}

SDSS~J0804 must be overluminous for its orbital period, relative to other AM CVn systems, and relative to theoretical models \citep{Nelemans2001}.  It is even possible for SDSS~J0804 that it has purely polar accretion, like the AM~Her class of cataclysmic variables.  The system was already known to have single-peaked optical emission lines \citep{Roelofs2009}, indicating that either the system is nearly face-on, or, more likely, that the accretion disk does not reach very close to the accreting white dwarf.  Additionally, the system shows a ``spur" in its phase resolved spectroscopic optical emission line profile \citep{Roelofs2009}, something also seen in some polar cataclysmic variables, which have no disks \citep{1987MNRAS.224..987R}.  { Following equation (2), the magnetic field would need to be about $10^5$ G to truncate the accretion disc outside the circularization radius.}

\section{Discussion}

\subsection{Magnetic braking and enhanced accretion}
The discovery that the accretor in Gaia14aae has a dynamically important magnetic field solves one of its core mysteries.  \cite{Campbell} had found that both its current luminosity, and the fact that it showed an outburst, to be indicators that the system is accreting at a higher rate than expected for an AM~CVn system of this orbital period under standard binary evolution assumptions.  { SDSS~0804 is similarly overluminous, and hence similarly is likely to be displaying enhanced accretion.}  

The accretion rates of AM~CVn systems can be enhanced by magnetic braking if the accretor's magnetic field is of order $10^3$G or more \citep{2010arXiv1006.4112F} and there is even a quite modest disk wind in the system (to first order, the angular momentum transport does not depend on the rate of mass loss in the wind \citep{2010arXiv1006.4112F}).  Disk winds are clearly present in outbursts of cataclysmic variables \citep{Cordova1982} and AM CVn systems \citep{Wade2007} and there is good evidence even in quiescence that the CVs have strong disk winds \citep{Hernandez2019}.  The mechanism by which the accretor's magnetic braking affects the binary orbital evolution is relevant only when the accretor is tidally locked to the orbital period, something which appears to be true for AM CVn systems, but not for standard cataclysmic variables in which there is an accretion disk \citep{2006MNRAS.371.1231R,2010arXiv1006.4112F}. 

\subsection{Implications of the magnetic braking scenario for angular momentum transport mechanisms in AM CVn systems}
The implications of the evidence for magnetic braking in AM CVn systems for their gravitational wave signatures are important, and may manifest themselves in two ways.  First, if the angular momentum transport is strongly affected by magnetic braking, then the period derivatives observed may be faster than those predicted from standard conservative mass transfer models with only gravitational radiation as a means of angular momentum transport.  As a result, standard templates used to detect gravitational waves may not have large enough period derivatives for a given orbital period, and the template bank to be used for LISA may need to be larger than currently presumed.  Second, if a large fraction of the AM~CVn harbor accretors with large magnetic fields, the space density of such systems, especially in the range of periods from about 20--80 minutes, may also be reduced by their faster evolution.  These effects are less likely to be important for the shortest period AM CVn systems, because the effects of the magnetic braking are much more weakly sensitive to orbital period than are the effects of gravitational radiation \citep{2010arXiv1006.4112F}.  On the other hand, benefits may exist as well.  For systems that are very well characterized in the gravitational wave band, with well-measured amplitudes that make predictions for the strength of the gravitational wave emission, it may be possible to compare the period derivatives that are measured with the ones expected from gravitational radiation alone, and to estimate the accreting white dwarf's magnetic fields from the LISA data.

The standard theory of AM~CVn evolution consists of systems moving to longer period as the donor white dwarf loses mass, and expands due to its degenerate nature.  The rate at which the mass loss takes place is set by finding the mass loss rate at which the angular momentum transport due to gravitational radiation yields an orbit that expands such that the donor star remains exactly in Roche lobe contact with the accretor.  Adding a new mechanism to transport angular momentum outside the binary will have two direct effects: (1) it will lead to more rapid orbital evolution of the system and (2) it will lead to more rapid mass transfer.

These two effects, in turn, mean that the period distribution of AM~CVn systems should be heavily skewed toward longer periods that predicted, and that the long period systems should be systematically brighter than expected.  Given the severe challenges in discovering the longest period AM~CVn, this may then help explain the apparent dearth of AM~CVn in observed samples \citep{vanroestel2022} relative to model predictions.  

Standard theory predicts that the systems at periods longer than about 45 minutes should be in ``stable low states", in which there are no outbursts \citep{Kotko2012}.  Gaia14aae was thus already enigmatic because of its outbursting behavior \citep{Campbell}.  Its outbursting behavior {\it is} consistent with its accretion rate inferred from both the temperature of the white dwarf and the X-ray luminosity in quiescence.  Notably, all AM~CVn systems with orbital periods longer than 49 minutes are systematically brighter than the predictions from standard models\citep{ramsay2018}, perhaps indicating that this magnetic braking effect is ubiquitous, and just harder to measure in the other long period systems due to lower quality X-ray data.

This, in turn, suggests that the accretion rate in both Gaia14aae must be higher than that from model tracks that include only gravitational radiation \citep{Deloye2007} by a factor of about 50.  If the solution is that magnetic braking dominates the angular momentum transport in this system, then the magnetic field should be about $10^4$ G, following figure 2 of \citet{2010arXiv1006.4112F}.  This is in good agreement with the finding of the double-peaked pulsations in the X-rays.  { If the accretion rate is overestimated by assuming the white dwarf temperature is reliable for estimating it, then significantly lower magnetic fields could be accommodated.

The single-peaked pulsations SDSS~J0804 imply a magnetic field of $\sim 10^5$ G.  A magnetic field of that size, which truncates the accretion disk relatively far from the surface of the white dwarf, can suppress the disk instability model by turning the region in which the hottest part of the accretion disk would exist into a region with accretion along the magnetic poles, that comes out as X-rays.  Dwarf nova outbursts are rare, possibly non-existent, in the intermediate polar class of cataclysmic variables for this reason \citep{2017A&A...602A.102H}, and completely absent in the polar class, in which no accretion disk forms.  

This would seem to be at odds with the results of \citet{2010arXiv1006.4112F}, which shows a dependence of approximately $\dot{M}\propto{B}^2$, so that an accretion rate about 100 times that of Gaia14aae would be expected, given the 10 times higher magnetic field. 
 In the context of a scenario where mass transfer is enhanced by magnetic braking, one possibility is that there is an accretion disc in SDSS~0804, and the magnetic field is well below $10^5$ G.  Another is that the effects of magnetic braking at a given magnetic field and the accretion rate in Gaia14aae have been overestimated by similar amounts.  A final possibility is that the magnetic braking process is fundamentally different in the absence of actual disc formation.}

\subsection{Tides?}
{ An alternative means of affecting the evolution of AM~CVn systems relative to the standard gravitational wave scenario is via tides.  This appears unlikely for two reasons: tides tend to reduce, rather than increase accretion rates \citep{2023ApJ...949...95B}, and tides are unlikely to lead to dramatic difference between systems at a given orbital period, especially for longer period systems that have had longer to synchronize \citep{2023ApJ...949...95B}.}

\subsection{Implications for long-term evolution of AM~CVn systems and their orbital period distribution}

Interestingly, if $\sim10^4$ G is a typical value for the magnetic fields in AM CVn, the magnetic braking would start to be the dominate source of angular momentum transport only for systems with periods longer than about 20 minutes.  It is most likely that there exists a broad range of magnetic fields in the AM~CVn systems just as there does for cataclysmic variables with hydrogen-rich donors.  Relatively few of these objects are the subjects of long observations with sensitive X-ray telescopes, and it is likely that performing such observations would reveal some more of these objects.

The effects of magnetic braking may be profound for both the orbital period distribution of the AM~CVn systems, and for the gravitational wave search approaches with LISA.  If the effect is important at periods as short as 20 minutes, it will affect some of the strongest mass-transferring gravitational wave sources for LISA.  The space densities of these systems will be reduced, as they will move through their orbital period evolution significantly faster than predicted by models with pure gravitational wave evolution.  

%\section{Results}\label{sec2}

% In setting up this template for *Science* papers, we've used both
% the \section* command and the \paragraph* command for topical
% divisions.  Which you use will of course depend on the type of paper
% you're writing.  Review Articles tend to have displayed headings, for
% which \section* is more appropriate; Research Articles, when they have
% formal topical divisions at all, tend to signal them with bold text
% that runs into the paragraph, for which \paragraph* is the right
% choice.  Either way, use the asterisk (*) modifier, as shown, to

\section*{Acknowledgments}
We thank Matthew Green for useful discussions, { and an anonymous referee for comments that improved the quality of this article}.  TK acknowledges support from the NSF through grant AST \#2107982, from NASA through grant 80NSSC22K0338 and from STScI through grant HST-GO-16659.002-A. Co-funded by the European Union (ERC, CompactBINARIES, 101078773). Views and opinions expressed are however those of the author(s) only and do not necessarily reflect those of the European Union or the European Research Council. Neither the European Union nor the granting authority can be held responsible for them.
\section{Data availability statement}
All new data presented in this Letter are accessible from the XMM-Newton archives.
\bibliography{scibib}

\begin{thebibliography}{}
\makeatletter
\relax
\def\mn@urlcharsother{\let\do\@makeother \do\$\do\&\do\#\do\^\do\_\do\%\do\~}
\def\mn@doi{\begingroup\mn@urlcharsother \@ifnextchar [ {\mn@doi@}
  {\mn@doi@[]}}
\def\mn@doi@[#1]#2{\def\@tempa{#1}\ifx\@tempa\@empty \href
  {http://dx.doi.org/#2} {doi:#2}\else \href {http://dx.doi.org/#2} {#1}\fi
  \endgroup}
\def\mn@eprint#1#2{\mn@eprint@#1:#2::\@nil}
\def\mn@eprint@arXiv#1{\href {http://arxiv.org/abs/#1} {{\tt arXiv:#1}}}
\def\mn@eprint@dblp#1{\href {http://dblp.uni-trier.de/rec/bibtex/#1.xml}
  {dblp:#1}}
\def\mn@eprint@#1:#2:#3:#4\@nil{\def\@tempa {#1}\def\@tempb {#2}\def\@tempc
  {#3}\ifx \@tempc \@empty \let \@tempc \@tempb \let \@tempb \@tempa \fi \ifx
  \@tempb \@empty \def\@tempb {arXiv}\fi \@ifundefined
  {mn@eprint@\@tempb}{\@tempb:\@tempc}{\expandafter \expandafter \csname
  mn@eprint@\@tempb\endcsname \expandafter{\@tempc}}}

\bibitem[\protect\citeauthoryear{{Amaro-Seoane} et~al.,}{{Amaro-Seoane}
  et~al.}{2022}]{LISA}
{Amaro-Seoane} P.,  et~al., 2022, \mn@doi [arXiv e-prints]
  {10.48550/arXiv.2203.06016}, {p. arXiv:2203.06016}

\bibitem[\protect\citeauthoryear{{Bahramian} et~al.,}{{Bahramian}
  et~al.}{2017}]{Bahramian2017}
{Bahramian} A.,  et~al., 2017, \mn@doi [\mnras] {10.1093/mnras/stx166}, 
{467, 2199}

\bibitem[\protect\citeauthoryear{{Bailer-Jones}, {Rybizki}, {Fouesneau},
  {Demleitner}  \& {Andrae}}{{Bailer-Jones} et~al.}{2021}]{BailerJones}
{Bailer-Jones} C.~A.~L.,  {Rybizki} J.,  {Fouesneau} M.,  {Demleitner} M.,
  {Andrae} R.,  2021, \mn@doi [\aj] {10.3847/1538-3881/abd806},
{161, 147}

\bibitem[\protect\citeauthoryear{{Begari} \& {Maccarone}}{{Begari} \&
  {Maccarone}}{2023}]{2023arXiv231206007B}
{Begari} T.,  {Maccarone} T.~J.,  2023, \mn@doi [JAAVSO, in press, arXiv
  e-prints] {10.48550/arXiv.2312.06007}, \href
  {https://ui.adsabs.harvard.edu/abs/2023arXiv231206007B} {p. arXiv:2312.06007}

\bibitem[\protect\citeauthoryear{{Biscoveanu}, {Kremer}  \&
  {Thrane}}{{Biscoveanu} et~al.}{2023}]{2023ApJ...949...95B}
{Biscoveanu} S.,  {Kremer} K.,   {Thrane} E.,  2023, \mn@doi [\apj]
  {10.3847/1538-4357/acc585},
{949, 95}

\bibitem[\protect\citeauthoryear{{Campbell} et~al.,}{{Campbell}
  et~al.}{2015}]{Campbell}
{Campbell} H.~C.,  et~al., 2015, \mn@doi [\mnras] {10.1093/mnras/stv1224},
{452, 1060}

\bibitem[\protect\citeauthoryear{{Cordova} \& {Mason}}{{Cordova} \&
  {Mason}}{1982}]{Cordova1982}
{Cordova} F.~A.,  {Mason} K.~O.,  1982, \mn@doi [\apj] {10.1086/160291}, 
{260, 716}

\bibitem[\protect\citeauthoryear{{Deloye}, {Taam}, {Winisdoerffer}  \&
  {Chabrier}}{{Deloye} et~al.}{2007}]{Deloye2007}
{Deloye} C.~J.,  {Taam} R.~E.,  {Winisdoerffer} C.,   {Chabrier} G.,  2007,
  \mn@doi [\mnras] {10.1111/j.1365-2966.2007.12262.x}, {381, 525}

\bibitem[\protect\citeauthoryear{{Farmer} \& {Roelofs}}{{Farmer} \&
  {Roelofs}}{2010}]{2010arXiv1006.4112F}
{Farmer} A.,  {Roelofs} G.,  2010, arXiv e-prints, \href
  {https://ui.adsabs.harvard.edu/abs/2010arXiv1006.4112F} {p. arXiv:1006.4112}

\bibitem[\protect\citeauthoryear{{Frank}, {King}  \& {Raine}}{{Frank}
  et~al.}{2002}]{2002apa..book.....F}
{Frank} J.,  {King} A.,   {Raine} D.~J.,  2002, {Accretion Power in
  Astrophysics: Third Edition}

\bibitem[\protect\citeauthoryear{{Gaia Collaboration} et~al.,}{{Gaia
  Collaboration} et~al.}{2021}]{GaiaEDR3}
{Gaia Collaboration} et~al., 2021, \mn@doi [\aap]
  {10.1051/0004-6361/202039657}, {649, A1}

\bibitem[\protect\citeauthoryear{{Hameury} \& {Lasota}}{{Hameury} \&
  {Lasota}}{2017}]{2017A&A...602A.102H}
{Hameury} J.~M.,  {Lasota} J.~P.,  2017, \mn@doi [\aap]
  {10.1051/0004-6361/201730760}, {602, A102}

\bibitem[\protect\citeauthoryear{{Hern{\'a}ndez Santisteban}
  et~al.,}{{Hern{\'a}ndez Santisteban} et~al.}{2019}]{Hernandez2019}
{Hern{\'a}ndez Santisteban} J.~V.,  et~al., 2019, \mn@doi [\mnras]
  {10.1093/mnras/stz798}, {486, 2631}

\bibitem[\protect\citeauthoryear{{Kotko}, {Lasota}, {Dubus}  \&
  {Hameury}}{{Kotko} et~al.}{2012}]{Kotko2012}
{Kotko} I.,  {Lasota} J.~P.,  {Dubus} G.,   {Hameury} J.~M.,  2012, \mn@doi
  [\aap] {10.1051/0004-6361/201219156}, {544, A13}

\bibitem[\protect\citeauthoryear{{Kraft}, {Mathews}  \& {Greenstein}}{{Kraft}
  et~al.}{1962}]{1962ApJ...136..312K}
{Kraft} R.~P.,  {Mathews} J.,   {Greenstein} J.~L.,  1962, \mn@doi [\apj]
  {10.1086/147381}, {136, 312}

\bibitem[\protect\citeauthoryear{{Kupfer}, {Steeghs}, {Groot}, {Marsh},
  {Nelemans}  \& {Roelofs}}{{Kupfer} et~al.}{2016}]{tidal_locking}
{Kupfer} T.,  {Steeghs} D.,  {Groot} P.~J.,  {Marsh} T.~R.,  {Nelemans} G.,
  {Roelofs} G.~H.~A.,  2016, \mn@doi [\mnras] {10.1093/mnras/stw126},{457, 1828}

\bibitem[\protect\citeauthoryear{{Nelemans}, {Portegies Zwart}, {Verbunt}  \&
  {Yungelson}}{{Nelemans} et~al.}{2001}]{Nelemans2001}
{Nelemans} G.,  {Portegies Zwart} S.~F.,  {Verbunt} F.,   {Yungelson} L.~R.,
  2001, \mn@doi [\aap] {10.1051/0004-6361:20010049}, {368, 939}

\bibitem[\protect\citeauthoryear{{Nelemans}, {Yungelson}, {van der Sluys}  \&
  {Tout}}{{Nelemans} et~al.}{2010}]{Nelemans2010}
{Nelemans} G.,  {Yungelson} L.~R.,  {van der Sluys} M.~V.,   {Tout} C.~A.,
  2010, \mn@doi [\mnras] {10.1111/j.1365-2966.2009.15731.x}, {401, 1347}

\bibitem[\protect\citeauthoryear{{Norton}, {Beardmore}, {Allan}  \&
  {Hellier}}{{Norton} et~al.}{1999}]{1999A&A...347..203N}
{Norton} A.~J.,  {Beardmore} A.~P.,  {Allan} A.,   {Hellier} C.,  1999, \aap, {347, 203}

\bibitem[\protect\citeauthoryear{{Pala} et~al.,}{{Pala} et~al.}{2020}]{Pala}
{Pala} A.~F.,  et~al., 2020, \mn@doi [\mnras] {10.1093/mnras/staa764}, {494, 3799}

\bibitem[\protect\citeauthoryear{{Postnov} \& {Yungelson}}{{Postnov} \&
  {Yungelson}}{2014}]{PostnovYungelson}
{Postnov} K.~A.,  {Yungelson} L.~R.,  2014, \mn@doi [Living Reviews in
  Relativity] {10.12942/lrr-2014-3}, {17, 3}

\bibitem[\protect\citeauthoryear{{Ramsay}, {Groot}, {Marsh}, {Nelemans},
  {Steeghs}  \& {Hakala}}{{Ramsay} et~al.}{2006}]{ramsay2006}
{Ramsay} G.,  {Groot} P.~J.,  {Marsh} T.,  {Nelemans} G.,  {Steeghs} D.,
  {Hakala} P.,  2006, \mn@doi [\aap] {10.1051/0004-6361:20065491}, {457, 623}

\bibitem[\protect\citeauthoryear{{Ramsay} et~al.,}{{Ramsay}
  et~al.}{2018}]{ramsay2018}
{Ramsay} G.,  et~al., 2018, \mn@doi [\aap] {10.1051/0004-6361/201834261}, {620, A141}

\bibitem[\protect\citeauthoryear{{Roelofs}, {Groot}, {Nelemans}, {Marsh}  \&
  {Steeghs}}{{Roelofs} et~al.}{2006}]{2006MNRAS.371.1231R}
{Roelofs} G.~H.~A.,  {Groot} P.~J.,  {Nelemans} G.,  {Marsh} T.~R.,   {Steeghs}
  D.,  2006, \mn@doi [\mnras] {10.1111/j.1365-2966.2006.10718.x}, {371, 1231}

\bibitem[\protect\citeauthoryear{{Roelofs} et~al.,}{{Roelofs}
  et~al.}{2009}]{Roelofs2009}
{Roelofs} G.~H.~A.,  et~al., 2009, \mn@doi [\mnras]
  {10.1111/j.1365-2966.2008.14288.x}, {394, 367}

\bibitem[\protect\citeauthoryear{{Rosen}, {Mason}  \& {Cordova}}{{Rosen}
  et~al.}{1987}]{1987MNRAS.224..987R}
{Rosen} S.~R.,  {Mason} K.~O.,   {Cordova} F.~A.,  1987, \mn@doi [\mnras]
  {10.1093/mnras/224.4.987}, {224, 987}

\bibitem[\protect\citeauthoryear{{Sokoloski} \& {Bildsten}}{{Sokoloski} \&
  {Bildsten}}{1999}]{Sokoloski}
{Sokoloski} J.~L.,  {Bildsten} L.,  1999, \mn@doi [\apj] {10.1086/307234},
{517, 919}

\bibitem[\protect\citeauthoryear{{Verbunt} \& {Zwaan}}{{Verbunt} \&
  {Zwaan}}{1981}]{1981A&A...100L...7V}
{Verbunt} F.,  {Zwaan} C.,  1981, \aap, {100, L7}

\bibitem[\protect\citeauthoryear{{Wade}, {Eracleous}  \& {Flohic}}{{Wade}
  et~al.}{2007}]{Wade2007}
{Wade} R.~A.,  {Eracleous} M.,   {Flohic} H. M.~L.~G.,  2007, \mn@doi [\aj]
  {10.1086/521649}, {134, 1740}

\bibitem[\protect\citeauthoryear{{van Roestel} et~al.,}{{van Roestel}
  et~al.}{2022}]{vanroestel2022}
{van Roestel} J.,  et~al., 2022, \mn@doi [\mnras] {10.1093/mnras/stab2421},{512, 5440}

\bibitem[\protect\citeauthoryear{{van Teeseling}, {Beuermann}  \&
  {Verbunt}}{{van Teeseling} et~al.}{1996}]{vanteesling}
{van Teeseling} A.,  {Beuermann} K.,   {Verbunt} F.,  1996, \aap, {315, 467}

\makeatother
\end{thebibliography}
\bibliographystyle{mnras}

%Here you should list the contents of your Supplementary Materials -- below is an example. 
%You should include a list of Supplementary figures, Tables, and any references that appear only in the SM. 
%Note that the reference numbering continues from the main text to the SM.
% In the example below, Refs. 4-10 were cited only in the SM.     

% For your review copy (i.e., the file you initially send in for
% evaluation), you can use the {figure} environment and the
% \includegraphics command to stream your figures into the text, placing
% all figures at the end.  For the final, revised manuscript for
% acceptance and production, however, PostScript or other graphics
% should not be streamed into your compliled file.  Instead, set
% captions as simple paragraphs (with a \noindent tag), setting them
% off from the rest of the text with a \clearpage as shown  below, and
% submit figures as separate files according to the Art Department's
% instructions.

\clearpage

%\noindent { Fig. 1.} Please do not use figure environments to set up your figures in the final (post-peer-review) draft, do not include graphics in yoursource code, and do not cite figures in the text using \LaTeX\ \verb+\ref+ commands.  Instead, simply refer to the figure numbers in the text per {\it Science\/} style, and include the list of captions at the end of the document, coded as ordinary paragraphs as shown in the \texttt{scifile.tex} template file.  Your actual figure files should be submitted separately.

\end{document}